\documentclass[prl,twocolumn,superscriptaddress,showpacs,floatfix,longbibliography]{revtex4-1}
\usepackage{mathrsfs,braket}
\usepackage{amssymb, amsbsy, amsmath, latexsym, dsfont, array, layout,
graphicx,mathrsfs,color,ulem,bm}
\usepackage[colorlinks=true,citecolor=blue,urlcolor=blue]{hyperref}

\newcommand{\one}{\mathds{1}}
\newcommand{\ketbrad}[1]{\left|{#1}\rangle\!\langle{#1}\right|}
\newcommand{\ketbra}[2]{\left|{#1}\rangle\!\langle{#2}\right|}

\begin{document}

\title{Experimental Simulation of Symmetry-Protected Higher-Order Exceptional Points with Single Photons}
\author{Kunkun Wang}
\affiliation{School of Physics and Optoelectronic Engineering, Anhui University, Hefei 230601, China}
\affiliation{Beijing Computational Science Research Center, Beijing 100084, China}
\author{Lei Xiao}
\affiliation{Beijing Computational Science Research Center, Beijing 100084, China}
\author{Haiqing Lin}
\affiliation{School of Physics, Zhejiang University, Hangzhou 310030, China}
\affiliation{Beijing Computational Science Research Center, Beijing 100084, China}
\author{Wei Yi}\email{wyiz@ustc.edu.cn}
\affiliation{Key Laboratory of Quantum Information, University of Science and Technology of China, CAS, Hefei 230026, China}
\affiliation{CAS Center for Excellence in Quantum Information and
Quantum Physics, University of Science and Technology of China, Hefei 230026,
China}
\author{Emil J. Bergholtz}\email{emil.bergholtz@fysik.su.se}
\affiliation{Department of Physics, Stockholm University, AlbaNova University Center, 106 91 Stockholm, Sweden}
\author{Peng Xue}\email{gnep.eux@gmail.com}
\affiliation{Beijing Computational Science Research Center, Beijing 100084, China}

\begin{abstract}
{\bf Exceptional points (EPs) of non-Hermitian (NH) systems have recently attracted increasing attention due to their rich phenomenology and intriguing applications. Compared to the predominantly studied second-order EPs, higher-order EPs have been assumed to play a much less prominent role because they generically require the tuning of more parameters. Here we experimentally simulate two-dimensional topological NH band structures using single-photon interferometry, and observe topologically stable third-order EPs obtained by tuning only two real parameters in the presence of symmetry. In particular, we explore how different symmetries stabilize qualitatively different third-order EPs: the parity-time symmetry leads to a generic cube-root dispersion, while a generalized chiral symmetry implies a square-root dispersion coexisting with a flat band. Additionally, we simulate fourfold degeneracies, composed of the non-defective twofold degeneracies and second-order EPs. Our work reveals the abundant and conceptually richer higher-order EPs protected by symmetries and offers a versatile platform for further research on topological NH systems.
}
\end{abstract}

\maketitle
Exceptional points (EPs) of non-Hermitian (NH) systems are branch point singularities in the parameter space, which emerge at the turning points of the dispersions of the interface states~\cite{MA19,ORN19,WLG19,CLLnp20}. EPs exhibit fascinating topological phenomena~\cite{DGH01,XMJ16,LHY21,ZLL21}, and lead to intriguing applications, such as sensing~\cite{HHW17,COZ17,LLS19,CLL20,W20,YMT20,PR22}, unidirectional wave propagation~\cite{LRE11,RBM12,YZ13,CJH14}, chiral laser emission~\cite{POL16}, laser linewidth broadening~\cite{ZPO18}, and laser mode selection~\cite{POR14,FWM14,HMH14,WSS21}.

The simplest case of EPs is the second-order EP, that is, two-fold degeneracy, which intuitively occurs in a two-dimensional (2D) system, but can be promoted to knotted exceptional lines in three dimensions~\cite{CSB19,YH19,CHW19,WXB21}.
Generally, an $n$th-order EP is stable in a $(2n-2)$-dimensional NH system~\cite{HRH20}, which makes them qualitatively more abundant than degeneracies in Hermitian systems. When exploring the NH systems with symmetries, the dimension for the occurrence of generic second-order EPs can be further reduced from two to one, leading to the observation of stable second-order EPs even in one-dimensional systems~\cite{BCK19,ZLL19,YPK19,OY19}.
Similarly, generic NH symmetries have been found to reduce the dimension for the occurrence of the third-order EPs from four to two~\cite{MB21,DYH21,SK22}.
Moreover, different symmetries may also entail qualitatively different phenomenology for EPs with the same order~\cite{MB21}.
Although higher-order EPs have been studied by a number of works theoretically, their experimental realization and direct observation appear to be rather difficult, as
ingenious designs are required to simulate the NH dynamics~\cite{TJD20,TDM21} and explore the band structures.

Here, by using single-photon interferometry, we overcome the difficulty of building NH systems with a large number of tunable parameters, and simulate a 2D NH system in the reciprocal space. With interferometric measurements, complex eigenenergies are measured directly, enabling us to construct the band structures of the 2D NH system. In particular, we experimentally observe and characterize two distinct types of symmetry-protected third-order EPs. We also experimentally confirm the stability of the third-order EPs with respect to perturbations. Because they are protected by either the parity-time (PT) or a generalized chiral P symmetry, these EPs disappear upon introducing symmetry-breaking perturbations. Our experimental results demonstrate that the energy near the third-order EP exhibits a generic $\thicksim k^{1/3}$ dispersion enforced by PT symmetry. The coexistent exceptional ring (ER) composed by the second-order EP bounds an open Fermi surface, which is also the boundary between the PT-unbroken and PT-broken regimes. By contrast, an anomalous $\thicksim k^{1/2}$ dispersion is observed away from the chiral-P-symmetry-protected third-order EPs.

\begin{figure*}
  \centering
\includegraphics[width=0.9\textwidth]{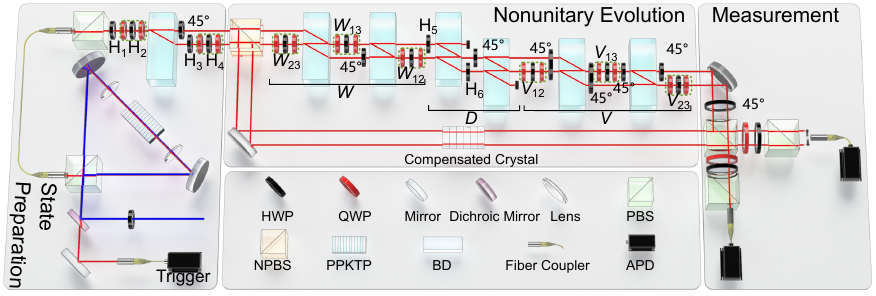}
\caption{{\bf Experimental setup.} State preparation is achieved by subjecting the signal photons to a polarizing beam splitter (PBS), five half-wave plates (HWPs), four quarter-wave plates (QWPs) and a beam displacer (BD). Single photons are generated by spontaneous parametric down-conversion (SPDC) in the periodically poled potassium titanyl phosphate (PPKTP) crystal. Two other spatial modes of photons are introduced after the photons pass through a non-polarizing beam splitter (NPBS). The transmitted photons experience a nonunitary evolution that is realized via the interferometric network. The interferometric network is composed by sets of wave plates and BDs. The reflected photons serve as references for interferometric measurements. The eigenenergies are encoded in the complex phase shift between the photons in different spatial modes, which are measured via interferometric measurements. The photons are detected by avalanche photodiodes (APDs).}
\label{fig:setup}
\end{figure*}

We further extend our experiment to simulate 2D NH models with four bands, where PT and P symmetries have radically different implications. For the PT symmetry, the third-order EPs and ER are present regardless of the additional bands. In contrast, for the P-symmetry case, the third-order EPs are forbidden by the additional bands. We also observe fourfold degeneracies, each composed by a non-defective twofold degeneracy and a second-order EP~\cite{SSR22,YSH21}. Our results thus experimentally expose the abundance of higher-order EPs whose codimensions are reduced in the presence of NH symmetries, thus offering potential applications in efficient device design. 

\begin{figure*}
  \centering
\includegraphics[width=0.9\textwidth]{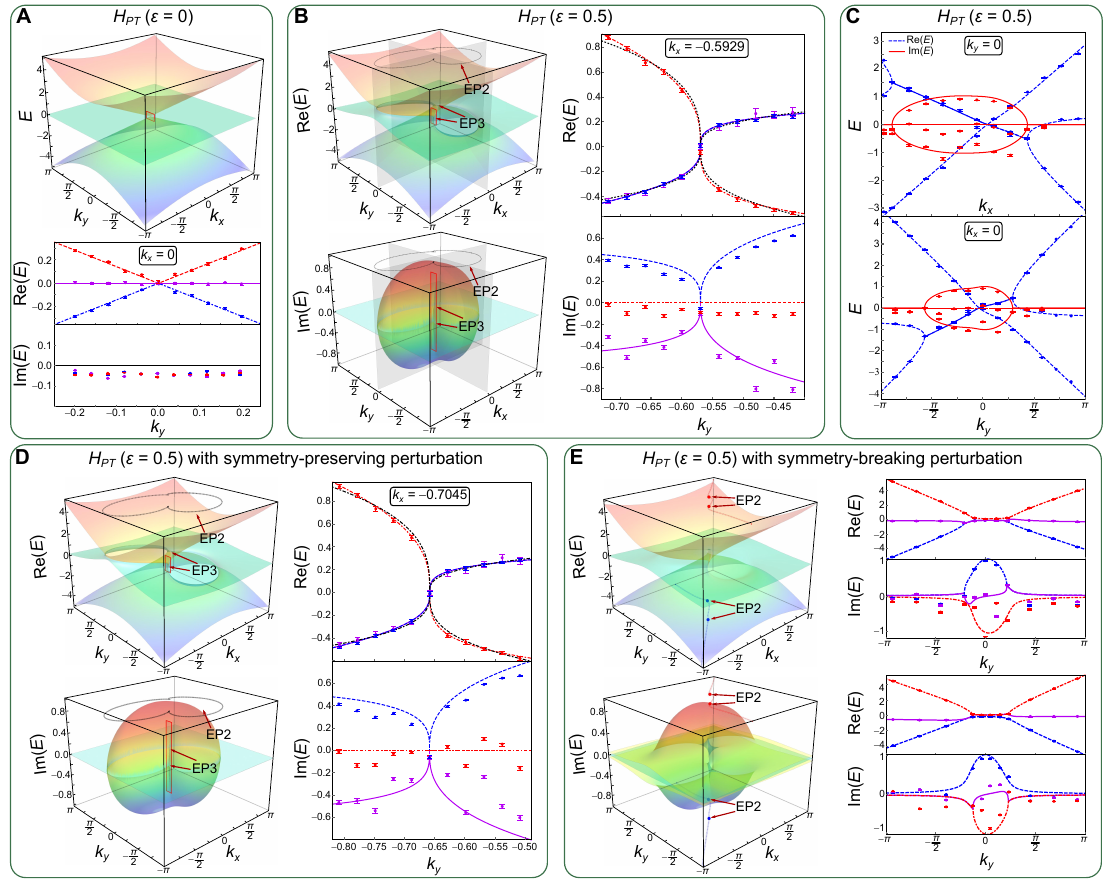}
\caption{{\bf Observation of the PT-symmetry-protected third-order EPs.} The real and imaginary parts of the eigenenergies of $H_{PT}$ with $\epsilon=0$ ({\bf A}) and $\epsilon=0.5$ ({\bf B} and {\bf C}), respectively, as functions of the momentum. The colored surfaces correspond to the theoretical results, where parameters for the experimental measurements are chosen within the range of red squares. Black dotted lines in the right columns of ({\bf B} and {\bf D}) correspond to the results fitted by $\thicksim k^{1/3}$. The measured real (blue) and imaginary (red) parts of the eigenenergies with $k_y=0$ and $k_x=0$ as functions of $k_x$ and $k_y$ are shown in the top and bottom panels of ({\bf C}). Effects due to the symmetry-preserving perturbation $i\pi(\lambda_4+\lambda_5)/20$ and the symmetry-broken perturbation $\sum_{i=1}^{8}\delta_i\lambda_i$ for $H_{PT}$ with $\epsilon=0.5$ are shown in ({\bf D} and {\bf E}), respectively, where $\delta_i\in[-\pi/20, \pi/20]$ are chosen randomly. The red and blue dotted lines in the left column of ({\bf E}) depict the generalized arc degeneracies, where the measured results along them are shown in the right column. Experimental data are represented by symbols and theoretical results by colored lines. Error bars are obtained by assuming Poisson statistics in the photon-number fluctuations, indicating the statistical uncertainty. EP2: the second-order EP; EP3: the third-order EP.}
\label{fig:data1}
\end{figure*}

\begin{figure*}
  \centering
\includegraphics[width=0.9\textwidth]{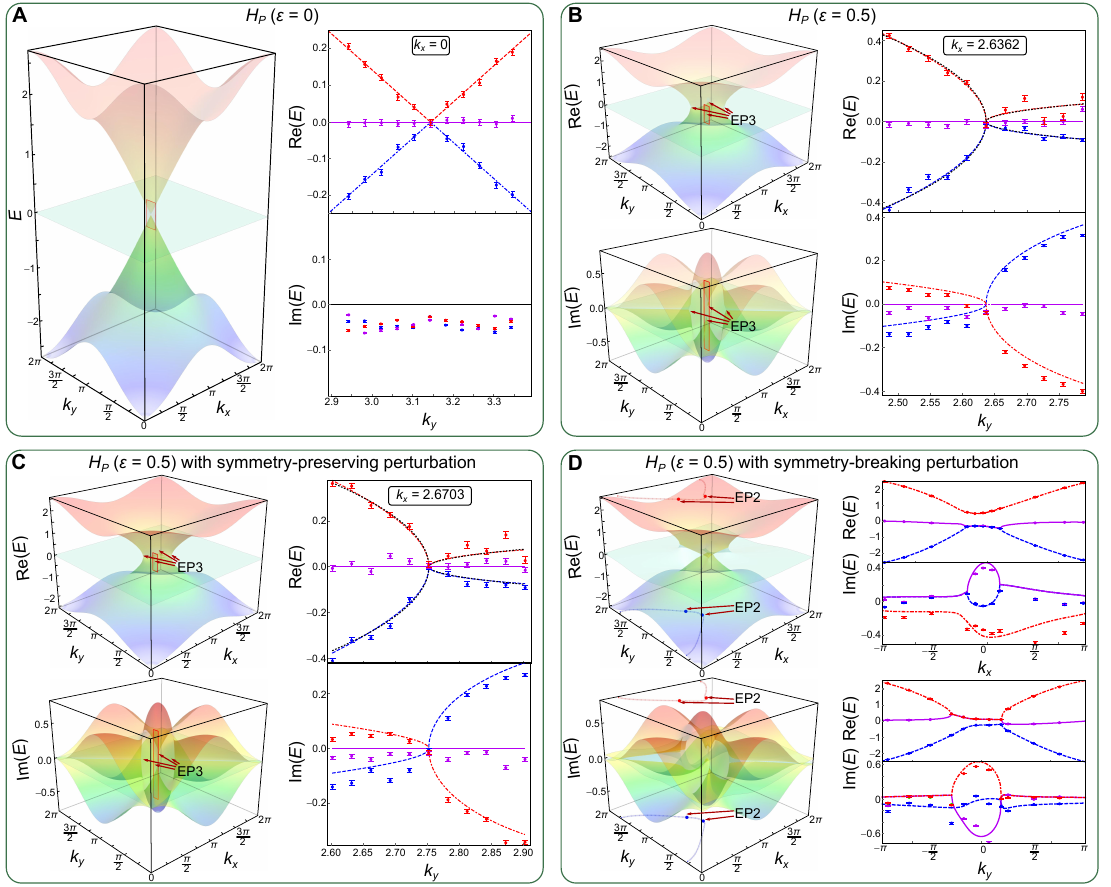}
\caption{{\bf Observation of the P-symmetry-protected third-order EPs.} The real and imaginary parts of the eigenenergies of $H_{P}$ with $\epsilon=0$ ({\bf A}) and $\epsilon=0.5$ ({\bf B}), respectively, as functions of the momentum. The colored surfaces correspond to the theoretical results, where parameters for the experimental measurements are chosen within the range of red squares. Black dotted lines in the right columns of ({\bf B}) and ({\bf C}) correspond to the results fitted by $\thicksim k^{1/2}$. Effects due to the symmetry-preserving perturbation $i\pi\lambda_1/20$ and the symmetry-broken perturbation $\sum_{i=1}^{8}\delta_i\lambda_i$ for $H_{P}$ with $\epsilon=0.5$ are shown in ({\bf C}) and ({\bf D}), respectively, where $\delta_i\in[-\pi/20, \pi/20]$ are chosen randomly.
The red and blue dotted lines in the left column of ({\bf D}) depict the generalized arc degeneracies, where the measured results along them are shown in the right column. Experimental data are represented by symbols and theoretical results by colored lines. Error bars are obtained by assuming Poisson statistics in the photon-number fluctuations, indicating the statistical uncertainty.}
\label{fig:data2}
\end{figure*}

\begin{figure*}
  \centering
\includegraphics[width=0.9\textwidth]{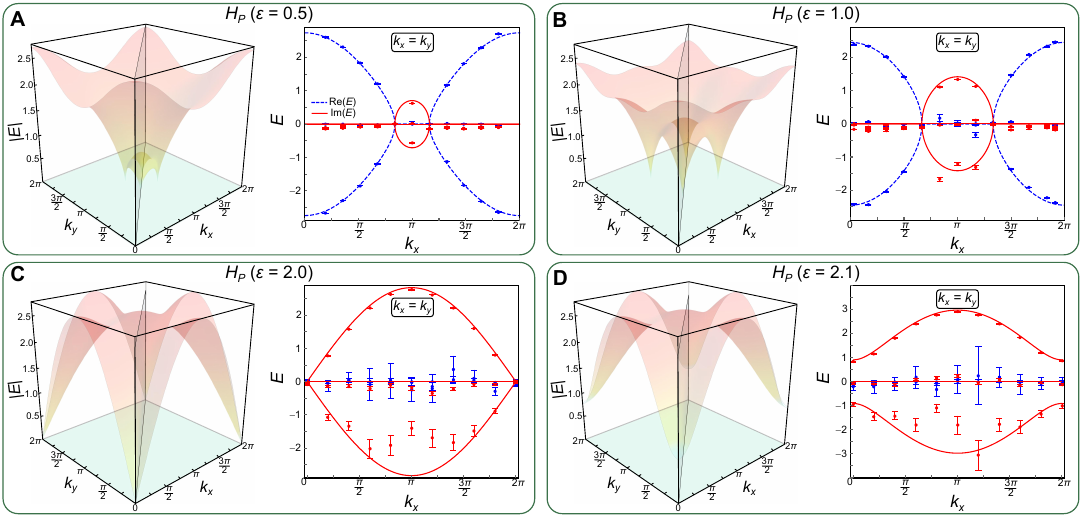}
\caption{{\bf Energy spectra for the P-symmetric three-band model.} The colored surfaces correspond to the absolute values for the eigenenergies of $H_P$ with $\epsilon=0.5$ ({\bf A}), $\epsilon=1.0$ ({\bf B}), $\epsilon=2.0$ ({\bf C}) and $\epsilon=2.1$ ({\bf D}), respectively. The gray planes correspond to the surfaces with $k_x=k_y$. Measured real (blue) and imaginary (red) results are represented by symbols and the corresponding theoretical ones by lines. Error bars are obtained by assuming Poisson statistics in the photon-number fluctuations, indicating the statistical uncertainty.}
\label{fig:data3}
\end{figure*}

{\bf Results}

{\bf Experimental setup.}
We explore the symmetry-protected third-order EPs by simulating the dynamics of the corresponding three-band NH Bloch Hamiltonians $H(\mathbf{k})$. The complex eigenenergies are measured via interferometric measurements~\cite{WXB21}.

As illustrated in Fig. 1, our experiment involves three stages: state preparation, nonunitary evolution, and measurement. The basis states of the three-band system are encoded into the hybrid polarization-spatial modes of single photons. In the preparation stage, the states of single photons are initialized in the eigenstate $\ket{\psi_j(\mathbf{k})}$ of $H(\mathbf{k})$ with the corresponding eigenenergy $E_j$. After passing through a non-polarizing beam splitter (NPBS), the transmitted photons go through the nonunitary evolution governed by $H(\mathbf{k})$ and the reflected photons remain unchanged. They interfere at the second PBS for interferometric measurements. To circumvent the quantum limit on achieving gain for single photons in the nonunitary-evolution stage, we map the time-evolution operator $U=e^{-iH(\mathbf{k})\tau}$ to a dissipative one $\tilde{U}=U/\sqrt{\Lambda}$. Here we fix the evolution time $\tau=1$ and $\hbar=1$, and take $\Lambda=\max_{j}|\zeta_{j}|$, where $\zeta_{j}$ is the eigenvalue of $e^{-iH}e^{iH^\dagger}$~\cite{WXB21,SMM18,WSX20}. The corresponding effective NH Hamiltonian for the mapped dissipative system is thus $\tilde{H}(\mathbf{k})=H(\mathbf{k})+i\ln\sqrt{1/\Lambda}\sigma_0$, where $\sigma_0$ is a $3\times3$ identity matrix. It follows that $\tilde{H}(\mathbf{k})$ and $H(\mathbf{k})$ have the same eigenstates, and their eigenenergies are related and satisfy $\tilde{E}_j=E_j-i\ln\sqrt{\Lambda}$. In the measurement stage, the complex inner products of $\xi=e^{-i\tilde{E}_j}=\bra{\psi_j}\tilde{U}\ket{\psi_j}$ are obtained through the interferometric measurements. The corresponding eigenenergies are then calculated from the experimentally measured $\xi$ (see Materials and Methods).


{\bf PT-symmetry protected third-order EPs.}
First, we consider a three-band linearized, higher-spin Dirac-like NH Hamiltonians with PT symmetry in a 2D reciprocal space~\cite{MB21}
\begin{align}\label{eq:HPT}
H_{PT}=&k_x\lambda_1+i\epsilon(\lambda_2+\lambda_4+\lambda_5)+(k_y-i\epsilon)\lambda_6\\ \nonumber
&-(k_y+i\epsilon)\lambda_7-\frac{\epsilon}{2}(\lambda_3+\sqrt{3}\lambda_8).
\end{align}
Here $\lambda_i$ ($i=1,2,\cdots,8$) denotes the Gell-Mann matrix~\cite{JAB08}. The model is Hermitian at $\epsilon=0$. Non-Hermiticity is introduced through finite $\epsilon$. Under the PT symmetry, $(PT)H^{\ast}_{PT}(PT)^{-1}=H_{PT}$ is satisfied, where $P=\text{diag}(1, -1, 1)$ and $T=\text{diag}(-1, 1, i)$ are unitary matrices associated with the parity and time-reversal operators~\cite{BL02} respectively, satisfying $T T^{\ast}=\pm1$ and $TP^{\ast}=PT$.

Figure 2 presents the real and imaginary parts of the eigenenergies of $H_{PT}$. At $\epsilon=0$, the Hamiltonian features a triple degeneracy, where two conical bands touch a flat band at $k_{x}=k_{y}=0$. As shown in Fig. 2A, by fixing $k_x=0$, we experimentally observe a linear dispersion near the degenerate point, consistent with the theoretical prediction~\cite{MB21}.

As illustrated in Fig. 2B, at $\epsilon=0.5$, the degenerate point splits into two third-order EPs, which occur at $\{k_x=-0.5929, k_y=-0.5694\}$ and $\{k_x=0.4701, k_y=0.6241\}$, respectively. Focusing on one of the third-order EPs, we fix $k_x=-0.5929$ and sample $11$ different values of $k_y$. The measured real and imaginary parts of the eigenenergies are shown in the right column of Fig. 2B. By fitting the power exponents with the formula $\thicksim k^\beta$, we confirm that the real eigenenergy exhibits a generic cube-root dispersion ($\thicksim k^{1/3}$) near the third-order EP.

Furthermore, an ER emerges in the momentum space, which is particularly interesting as the ER signals the PT transitions, and bounds an open Fermi surface~\cite{MB21}. Because the ER is a collection of second-order EPs,
we reveal its existence in Fig. 2C, where we measure the eigenenergies of $H_{PT}$ along the lines of $k_x=0$ and $k_y=0$, respectively. The second-order EPs are observed as the bifurcations of the real eigenenergies.
In between the two EPs, the imaginary parts of all the eigenenergies are nonzero, indicating the PT-symmetry broken region.
Therein, two of the measured eigenergies are approximately equal, indicating the emergence of an open Fermi surface.


As a feature of symmetry protection, the existence of the third-order EPs is robust against symmetry-preserving perturbations. As shown in Fig. 2D, under a small PT-symmetry-preserving perturbation $i\pi(\lambda_4+\lambda_5)/20$, both third-order EPs persist, but are shifted in parameter space. By contrast, the EPs disappear as illustrated in Fig. 2E, by introducing a general, symmetry-breaking perturbation  $\sum_{i=1}^{8}\delta_i\lambda_i$ ($\delta_i\in[-\pi/20, \pi/20]$). The perturbed spectrum exhibits branch cuts that are terminated by paired second-order EPs~\cite{YSH21}. These paired second-order EPs further reveal a transition from regions where the real parts of the eigenenergies are degenerate, to those with degeneracy in the imaginary parts.

{\bf P-symmetry protected third-order EPs.}
To demonstrate the features of NH model with P symmetry, we consider a NH Lieb lattice~\cite{MB21,L89}
\begin{align}\label{HP}
H_{P}=&(1+\cos k_x-i\epsilon)\lambda_1+(1+\cos k_y+i\epsilon)\lambda_6\\ \nonumber
&-\sin k_x \lambda_2-\sin k_y \lambda_7,
\end{align}
which satisfies the relation $H_{P}=-PH_{P}P^{-1}$. Note that we consider the case where this symmetry acts locally in momentum space. For $\epsilon=0$, $H_{P}$ corresponds to the Hermitian Lieb-lattice model. In the momentum space, similar to the case of $H_{PT}$, a triple degeneracy of the eigenenergy exists at $k_x = k_y = \pi$, from which emerges two dispersive bands with linear scaling [see Fig. 3A]. We introduce the NH term by setting $\epsilon=0.5$. The triple degeneracy then splits into four third-order EPs that locate at $(k_x=\pm2.6362, k_y=\pm2.6362)$. Focusing on one of the EPs, we fix $k_x=2.6362$ and sample $11$ different $k_y$. In Fig. 3B, we show the measured real and imaginary components of the eigenenergies, where the third-order EP is visible near $k_y= 2.6362$. Different from the PT-symmetry-protected case, the two nonzero real bands exhibit an anomalous square-root dispersion near the third-order EP, that is, $\thicksim k^{1/2}$.  This is a typical behavior associated with the second-order EPs. As shown in Fig. 3C, we introduce a P-symmetry-preserving perturbation in the form of $i\pi\lambda_1/20$. The four third-order EPs persist, with a small shift in the parameter space. Similar to $H_{PT}$, third-order EPs are destroyed by the general, symmetry-breaking perturbation $\sum_{i=1}^{8}\delta_i\lambda_i$, leading to a branch cut terminated by a pair of second-order EPs (see Fig. 3D). 

Accompanying the P-symmetry-protected third-order EPs of $H_P$, there are arc-like degeneracies along the lines of $k_x=\pm k_y$, which correspond to the real and imaginary Fermi arcs with Re[$E_\pm$]=0 and Im[$E_\pm$]=0, respectively~\cite{MB21}. Experimentally, we again sample $11$ different parameters along the line of $k_x=k_y$ for $H_P$ with $\epsilon=0.5$. In Fig. 4A, we show the measured real and imaginary components of the bands, where both the real and imaginary Fermi arcs can be observed. The boundaries between them give the locations of the third-order EPs.
With increasing $\epsilon$, as illustrated in Figs. 4B and C, the third-order EPs move from the center of the Brillouin zone near $(k_x=\pi, k_y=\pi)$ to the edges.
At $\epsilon=2.0$, the imaginary Fermi arcs give way to real Fermi arcs, as pairs of third-order EPs recombine, giving rise to linear dispersions at the EPs.
Further increasing $\epsilon$ beyond $2$, as shown in Fig. 4D with $\epsilon=2.1$, the third-order EPs completely disappear, as the spectrum opens up complex gaps. The real Fermi arcs develop into closed line-degeneracies with Re[$E_\pm$]=0, slicing through the entire Brillouin zone.

\begin{figure*}
  \centering
\includegraphics[width=\textwidth]{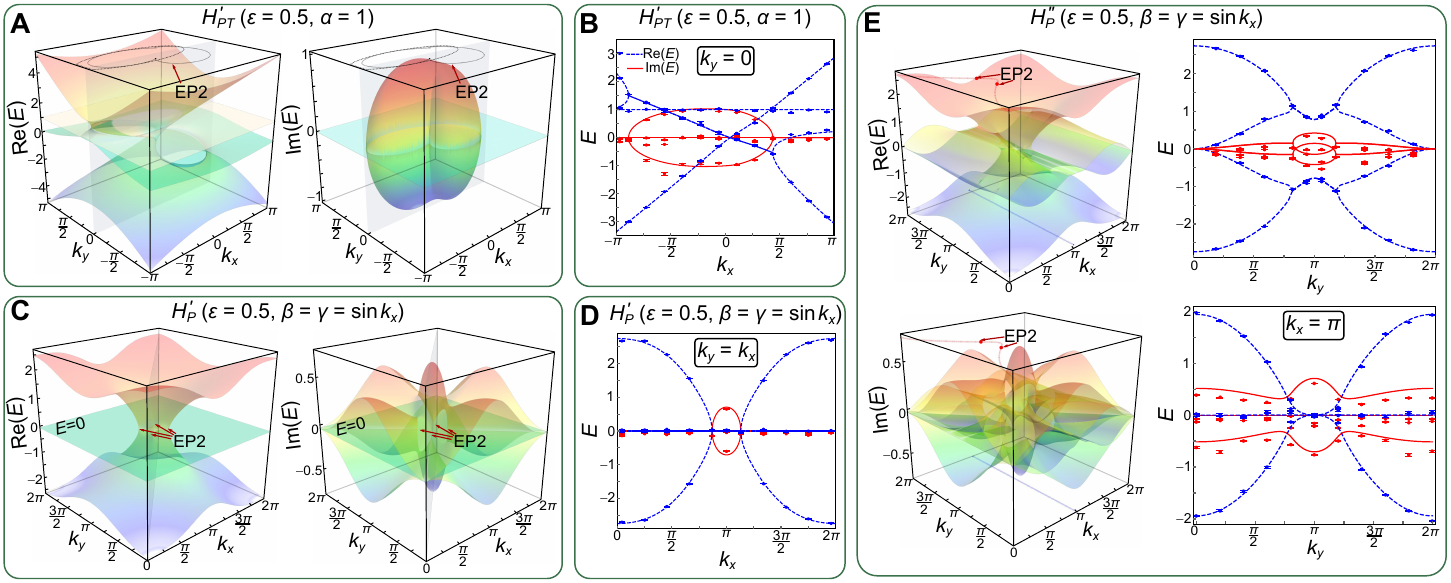}
\caption{{\bf Energy spectra for four-band models.} The colored surfaces correspond to the theoretical real and imaginary energy spectra of $H^\prime_{PT}$ with $\alpha=1$ ({\bf A}), $H^\prime_{P}$ with $\beta=\gamma=\sin k_x$ ({\bf C}) and $H^{\prime\prime}_{P}$ with $\beta=\gamma=\sin k_x$ ({\bf E}) by setting $\epsilon=0.5$. The gray planes correspond to the surfaces with $k_y=0$ ({\bf A}) and $k_y=k_x$ ({\bf C}), respectively, as chosen in our experiment. Measured real (blue) and imaginary (red) results are represented by points and the corresponding theoretical ones by lines for $H^\prime_{PT}$ ({\bf B}) and for $H^\prime_{P}$ ({\bf D}), respectively. The red dotted lines in the left column of ({\bf E}) depicts the generalized arc-like degeneracies. The blue dotted line corresponds to the non-defective two-fold degenerate line with $k_x=\pi$. The lines in the right column correspond to the theoretical results along the red dotted line (top) and $k_x=\pi$ (bottom), respectively. Measured results are represented by symbols. Error bars are obtained by assuming Poisson statistics in the photon-number fluctuations, indicating the statistical uncertainty.}
\label{fig:data4}
\end{figure*}

{\bf Generalization to symmetry-protected four-band models.}
We extend the system to the symmetry-protected four-band model governed by the Hamiltonian
\begin{align}
H^\prime_{PT}=&k_x\Gamma_1+i\epsilon(\Gamma_2+\Gamma_4+\Gamma_5)+(k_y-i\epsilon)\Gamma_6\\ \nonumber
&-(k_y+i\epsilon)\Gamma_7-\frac{\epsilon}{2}(\Gamma_3+\sqrt{3}\Gamma_8)+\frac{\Gamma_0-\sqrt{6}\Gamma_{15}}{4},
\label{eq:HPT4}
\end{align}
which is PT symmetric, with the symmetry operators $P^\prime=\text{diag}(1, -1, 1, 1)$ and $T^\prime=\text{diag}(-1,$ $1,i, 1)$. Here $\Gamma_0$ is a $4\times4$ identity matrix and $\Gamma_i$ ($i=1,2,\cdots,15$) denote the Gell-Mann matrices that span the Lie algebra of the SU(4) group.

Compared to $H_{PT}$ in Eq. 1, the eigenspectrum of $H^\prime_{PT}$ possesses an additional flat band at $E=\alpha$. As shown in Fig. 5A, third-order EPs persist under $\alpha=1$ and $\epsilon=0.5$, at the same locations as those of $H_{PT}$. A manifest difference is the emergence of an additional ER composed by the second-order EPs in the PT-symmetric four-band model. In Fig. 5B,  by sampling the momentum of $k_x$ along the line of $k_y=0$, we experimentally confirm the existence of the extra real eigenenergy and the additional ER.

For P-symmetry-protected four band models, the restricted NH Hamiltonian becomes
\begin{align}
H^{\prime}_{P}=&(1+\cos k_x-i\epsilon)\Gamma_1+(1+\cos k_y+i\epsilon)\Gamma_6-\sin k_x \Gamma_2\\ \nonumber
&-\sin k_y\Gamma_7+\frac{(\beta+\gamma)}{2}\Gamma_{11}+\frac{i(\beta-\gamma)}{2}\Gamma_{12},
\label{eq:HP4}
\end{align}
where $\beta$ and $\gamma$ are arbitrary complex numbers.
Here Hamiltonian $H^{\prime}_{P}$ satisfies the relation $\text{det}[H^\prime_{P}]=\text{Tr}(H^{\prime}_{P})=\text{Tr}(H^{\prime3}_{P})=0$. Its corresponding eigenenergies are $\{0, 0, E_\pm^\prime\}$~\cite{SK22} with $E_\pm^\prime=\pm\sqrt{4+\beta\gamma-2\epsilon^2+(2-2i\epsilon)\cos k_x+(2+2i\epsilon)\cos k_y}$, where the third-order EPs are destroyed by the additional band. The two degenerate flat bands at zero energy correspond to the non-defective degeneracies with two distinct eigenstates~\cite{SSR22,YSH21}. Setting $\epsilon=0.5$ and tuning $k_x$ and $k_y$, we confirm that the two dispersive bands $E_\pm^\prime$ touch each other at $E_\pm^\prime=0$, to form the second-order EPs, where the local degeneracies become four-fold (see Figs. 5C and D).

Such four-fold degeneracies are not protected by the P symmetry. Changing the P-symmetry operator to $P^{''}=\text{diag}(1, -1, 1, -1)$~\cite{MB21}, we choose the non-Hermitian Hamiltonian in the form
\begin{align}
H^{\prime\prime}_{P}=&(1+\cos k_x-i\epsilon)\Gamma_1+(1+\cos k_y+i\epsilon)\Gamma_6-\sin k_x \Gamma_2\\ \nonumber
&-\sin k_y\Gamma_7 +\frac{(\beta+\gamma)}{2}\Gamma_{13}+\frac{i(\beta-\gamma)}{2}\Gamma_{14}.
\label{eq:HP4a}
\end{align}
As shown in Fig. 5E, the four-fold degeneracies are gapped out, with $\beta =
\gamma = \sin k_x$ and $\epsilon = 0.5$.

{\bf Conclusion}

By experimentally studying the symmetry-protected higher-order EPs of a series of 2D NH models, our work provides the first experimental confirmation that symmetries qualitatively enrich the phenomenology of higher-order degeneracies in the NH systems.
By revealing the abundance of higher-order EPs whose codimensions are reduced in the presence of NH symmetries, our results have rich implications for efficient device design.

EPs are generic features of the effective description of a vast range of complex systems ranging from mechanical systems to strongly interacting quantum materials. In contrast, our simple experimental scheme is highly controllable, scalable, and can be readily extended to the detailed study of NH models with arbitrary design and physical origin. It thus constitutes a versatile and efficient platform for the systematic exploration of eigenspectrum and dynamical properties in NH settings.

{\bf Materials and Methods}

{\bf Initial state preparation.}
As illustrated in Fig. 1, our experimental setup can be used to prepare arbitrary pure qutrit states. The basis states of the three-band systems are encoded by the polarizations and spatial modes of single photons, with $\ket{0}\Leftrightarrow\ket{H_1}$, $\ket{1}\Leftrightarrow\ket{H_2}$, and $\ket{2}\Leftrightarrow\ket{V_2}$. Here the subscripts denote the different spatial modes and $H$ ($V$) denotes the horizontal (vertical) polarization of single photons. To prepare the generic qutrit state $a\ket{H_1}+be^{i\varphi_1}\ket{H_2}+ce^{i\varphi_2}\ket{V_2}$, the photons are first initialized to the horizontal polarization by passing them through a PBS. The two spatial modes of photons are introduced after the photons pass through a beam displacer. The real coefficients $\{a, b, c\}$ with $a^2+b^2+c^2=1$, can be adjusted by controlling HWPs with the setting angles of H$_1=(\arcsin a)/2$ and H$_3=(\arctan c/b)/2$. As for the relative phases $\{\varphi_1, \varphi_2\}$ ranging from $0$ to $2\pi$, they can be adjusted by the sandwich-type set of HWP and QWPs, denoted as QWP-HWP-QWP. Setting the angles of QWPs at $45^\circ$ and HWP at $45^\circ-180^\circ\varphi/\pi$, one can achieve the phase operation of diag($e^{2i\varphi}, e^{-2i\varphi}$). Thus, by setting H$_2=45^\circ(\varphi_1+\varphi_2)/\pi+56.25^\circ$ and H$_4=45^\circ(\varphi_2-\varphi_1)+11.25^\circ$, we can tune the relative phases accordingly.

It follows that the initial state can be exactly prepared in one of the eigenstates of the NH Hamiltonian with prior information. Even if the eigenstates are unknown, we can still generate them as initial states by maximizing the probability of $P=|\bra{\Psi}\tilde{U}\ket{\Psi}|^2/\bra{\Psi}\tilde{U}^\dagger\tilde{U}\ket{\Psi}$. If and only if $\ket{\Psi}$ is one of the eigenstates of the NH Hamiltonian with its dissipative nonunitary unit-time evolution of $\tilde{U}$, $P$ is maximized to $1$. Thus, tuning the angles of the wave-plates in state preparation to maximize the measured $P$, the prepared state must tend to one of the target eigenstates (see section S3 and the Supplementary Materials for more details)).

{\bf Experimental implementation of $\tilde{U}$.}
To implement the nonunitary operation $\tilde{U}$, we first decompose it into $\tilde{U}=VD_3W$ using singular value decomposition~\cite{TRS18}. The operations of $V$ and $W$ are unitary and $D_3=\ketbrad{H_1}+\mu\ketbrad{H_2}+\nu\ketbrad{V_2}$ with $0\leq\mu,\nu\leq 1$. We further decompose the two unitary matrices $W$ and $V$ by the established methods in~\cite{WKZ17,RZB94}, i.e., $W=W_{23}W_{13}W_{12}$ and $V=V_{12}V_{13}V_{23}$, where $W_{ij}$ and $V_{ij}$ are the unitary operations acting on the 2D subspaces of the qutrit system, with the complementary subspace unchanged.
As shown in Fig. 1, the unitary operations of $W_{ij}$ and $V_{ij}$ are realized by recombining the photons into the certain spatial mode depending on their polarizations and applying a $2\times2$ transformation via waveplates. The diagonal matrix $D_3$ can be realized by introducing the mode-selective losses of photons, where the photon losses can be controlled by the HWPs with the setting angles of H$_5=(\arccos \nu)/2$ and H$_6=(\arccos \mu)/2$.

{\bf Interferometric measurements.}
As shown in Fig. 1, the single photons are generated by spontaneous parametric down-conversion, and prepared in one of the eigenstates $\ket{\psi_j}$ of $H(\mathbf{k})$ with the corresponding eigenenergies of $E_j$.  After photons pass through the NPBS, their state is $\left(\ket{t}\ket{\psi_j} + \ket{r}\ket{\psi_j} \right)/\sqrt{2}$, where $t$ and $r$ denote the transmitted and reflected modes of the single photons, respectively. Applying the nonunitary operation governed by $\tilde{H}(\mathbf{k})$ on the photons in $t$, the state evolves to $\left(e^{-i\tilde{E}_j}\ket{t}\ket{\psi_j} + \ket{r}\ket{\psi_j} \right)/\sqrt{2}$. After this stage, the interferometric measurements are performed to obtain the overlap between the states of the photons in the transmitted and reflected modes, and thus to extract the complex phase shift of $\tilde{E}_j$.

In the measurement stage, an HWP at $45^\circ$ is first applied on the polarization of the photons in the transmitted mode. After the photons pass through a PBS, their state evolves into
\begin{align*}
&\frac{1}{\sqrt{2}}\left(\alpha_j\ket{t_1^\prime}+\beta_j\ket{t_2^\prime}\right)\left(\ket{H}+e^{-i\tilde{E}_j}\ket{V}\right)
\\ \nonumber
&+\frac{1}{\sqrt{2}}\gamma_j\ket{r_2^\prime}\left(e^{-i\tilde{E}_j}\ket{H}+\ket{V}\right)
\end{align*}
with the assumption $\ket{\psi_j}=\alpha_j\ket{H_1}+\beta_j\ket{H_2}+\gamma_j\ket{V_2}$. Here $t_{1(2)}^\prime$ ($r_{2}^\prime$) denotes the transmitted (reflected) mode of the photons which are reflected by the NPBS after passing through the second PBS. Projective measurements are performed on the polarization of the photons in different spatial modes with the bases of $\{\ket{\pm}=\left(\ket{H}\pm\ket{V}\right)/\sqrt{2}, \ket{R}=\left(\ket{H}-i\ket{V}\right)/\sqrt{2}\}$. The coincidence counts are denoted as $\{N^{\pm}_{i}, N^{R}_{i}\}$, where $i\in\{t_1^\prime, t_2^\prime, r_2^\prime\}$ corresponds to the spatial mode. Then, we have
\begin{align*}
\xi=&\frac{\sum_i\left(N^+_{i}-N^-_{i}\right)+i\sum_i\left(N^+_{i}+N^-_{i}-2N^R_{i}\right)}{N_\text{tot}}
\\ \nonumber
&-\frac{2i\left(N^+_{r_2^\prime}+N^-_{r_2^\prime}-2N^R_{r_2^\prime}\right)}{N_\text{tot}} \nonumber,
\end{align*}
where $N_\text{tot}$ denotes the number of the input photons to achieve the state preparation.

Note that our setup can be readily extended to simulate the dynamic evolution and construct the energy spectrum of arbitrary NH Hamiltonians, by taking advantage of the extendable degrees of freedom of photons and specially designed interferometric network (see section S5 and the Supplementary Materials for more details).

\noindent{\bf Acknowledgements}

\noindent{\bf Funding:} This work was supported by the National Natural Science Foundation of China (NSFC with grant nos. 92265209 and 12025401). K.W. acknowledges support from NSFC (grant no. 12104009). L.X. acknowledges support from NSFC (grant no. 12104036). H.L. acknowledges support from NSFC (grant no. 12088101). W.Y. acknowledges support from NSFC (grant no. 11974331). E.J.B. acknowledges support from the Swedish Research Council (grant no. 2018-00313), the Knut and Alice Wallenberg Foundation (KAW) (grant nos. 2018.0460 and2019.0068), and the G\"{o}ran Gustafsson Foundation for Research in Natural Sciences and Medicine.

\noindent{\bf Author contributions:}
K.W. performed the experiments with contributions from L.X. P.X. designed the experiments, analyzed the results, and wrote part of the paper. E.J.B. and W.Y.
developed the theoretical aspects and revised the paper. H.L. participated in the discussion and revised the paper.

\noindent{\bf Competing interests:}
The authors declare that they have no competing interests.

\noindent{\bf Data and materials availability:}
All data needed to evaluate the conclusions in the paper are present in the paper and/or the Supplementary Materials.

\clearpage
\appendix

\pagebreak
\widetext
\begin{center}
\textbf{\large  Supplementary Materials for  ``Experimental Simulation of Symmetry-Protected Higher-Order Exceptional Points with Single Photons''}
\end{center}
\setcounter{equation}{0}
\setcounter{figure}{0}
\setcounter{table}{0}
\makeatletter
\renewcommand{\theequation}{S\arabic{equation}}
\renewcommand{\thefigure}{S\arabic{figure}}
\renewcommand{\bibnumfmt}[1]{[S#1]}

\subsection{S1 PT-symmetry-protected third-order exceptional points.}
In our experiment, we first consider the three-band linearized, higher-spin Dirac-like non-Hermitian (NH) Hamiltonians with parity-time (PT) symmetry in two-dimensional (2D) reciprocal space~\cite{MB21},
\begin{equation}
H_{PT}=k_x\lambda_1+i\epsilon(\lambda_2+\lambda_4+\lambda_5)+(k_y-i\epsilon)\lambda_6-(k_y+i\epsilon)\lambda_7-\frac{\epsilon(\lambda_3+\sqrt{3}\lambda_8)}{2}.
\end{equation}
Here $\lambda_i$ with $i=1,2,\cdots,8$ denote the Gell-Mann matrices~\cite{JAB08}
, i.e.,
\begin{align}
\lambda_1&=\begin{pmatrix}
0 & 1 & 0 \\
1 & 0 & 0 \\
0 & 0 & 0 \\
\end{pmatrix},
\lambda_2=\begin{pmatrix}
0 & -i & 0 \\
i & 0 & 0 \\
0 & 0 & 0 \\
\end{pmatrix},
\lambda_3=\begin{pmatrix}
1 & 0 & 0 \\
0 & -1 & 0 \\
0 & 0 & 0 \\
\end{pmatrix},
\lambda_4=\begin{pmatrix}
0 & 0 & 1 \\
0 & 0 & 0 \\
1 & 0 & 0 \\
\end{pmatrix},\\ \nonumber
\lambda_5&=\begin{pmatrix}
0 & 0 & -i \\
0 & 0 & 0 \\
i & 0 & 0 \\
\end{pmatrix},
\lambda_6=\begin{pmatrix}
0 & 0 & 0 \\
0 & 0 & 1 \\
0 & 1 & 0 \\
\end{pmatrix},
\lambda_7=\begin{pmatrix}
0 & 0 & 0 \\
0 & 0 & -i \\
0 & i & 0 \\
\end{pmatrix},
\lambda_8=\begin{pmatrix}
\frac{1}{\sqrt{3}} & 0 & 0 \\
0 & \frac{1}{\sqrt{3}} & 0 \\
0 & 0 & \frac{-2}{\sqrt{3}} \\
\end{pmatrix}.
\label{eq:GM}
\end{align}
For the spectrum of $H_{PT}$, all the three eigenenergies can be uniquely defined as $E_1=\alpha_++\alpha_-$, $E_2=\omega\alpha_++\omega^\ast\alpha_-$, and $E_3=\omega^\ast\alpha_++\omega\alpha_-$, where $\omega=(-1+\sqrt{3}i)/2$ and $\alpha_{\pm}\equiv\sqrt[3]{q\pm\sqrt{p^3+q^2}}$ with $p=(-k_x^2-2k_y^2+4\epsilon^2)/3$ and $q=-\epsilon[k_x^2-2k_y^2+4(k_y-k_x)\epsilon+\epsilon^2]/2$. The third-order EPs occur when $p=q=0$ is taken. The exceptional ring (ER) is composed by second-order EPs and exists on the curve of $p^3+q^2=0$, which specifies the PT transitions, with all the eigenvalues being real only in the exact PT region with $p^3+q^2 > 0$. For $p^3+q^2 < 0$, PT-symmetry is broken with $E_2=E_3^\ast$.

\begin{figure*}[b!]
\centering
\includegraphics[width=0.9\textwidth]{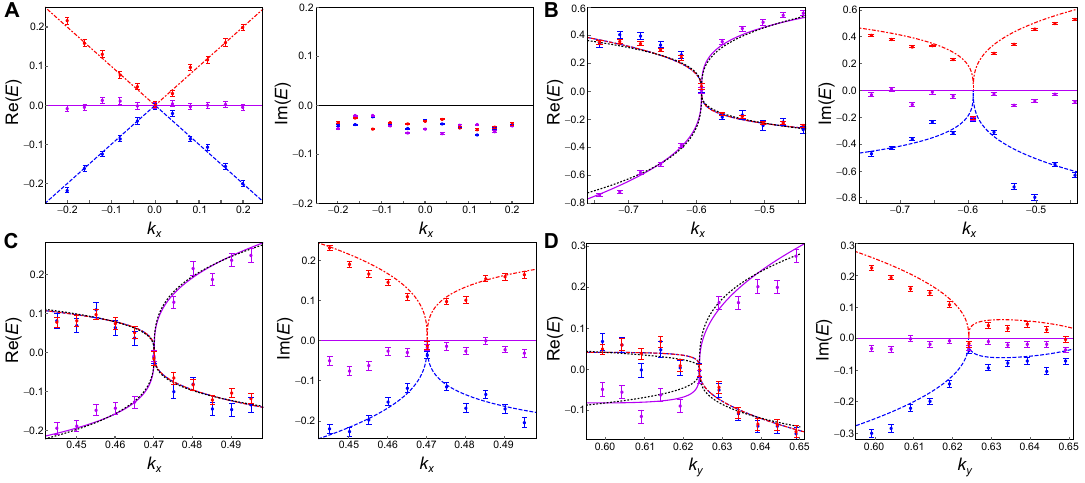}
\caption{{\bf Energy spectra for the PT-symmetric three-band model.} Measured (dots) and the corresponding theoretical (lines) eigenenergies of the PT-symmetric Hamiltonian $H_{PT}$ with fixed $(k_y=0, \epsilon=0)$ for ({\bf A}) and $(k_y=-0.5694, \epsilon=0.5)$ for ({\bf B}) as a function of the momentum $k_x$. The energy dispersion near the third-order EP at $(k_x=0.4701, k_y=0.6241)$ are observed by fixing $k_y=0.6241$ along $k_x$ in ({\bf C}), and fixing $k_x=0.4701$ along $k_y$ in ({\bf D}). The black dotted lines in the left column of ({\bf B}), ({\bf C}) and ({\bf D}) correspond to the results fitted by $\thicksim k^{1/3}$. Error bars are obtained by assuming Poisson statistics in the photon-number fluctuations, indicating the statistical uncertainty.}
\label{fig:dataS1}
\end{figure*}

In Figs. 2A and B of the main text, we characterize the energy dispersion along $k_y$ for $H_{PT}$ with $\epsilon=0$ and $\epsilon=0.5$. As compensation, we fix $k_y=0$ with $\epsilon=0$ and $k_y=-0.5694$ with $\epsilon=0.5$ for $H_{PT}$ with various $k_x$ experimentally. As shown Fig. S1A, the measured energy spectrum for the Hermitian model with $\epsilon=0$ presents a triple degeneracy with the same linear dispersion along $k_x$. The PT-symmetric NH model with $\epsilon=0.5$ has the energy spectrum which exhibits a cube-root dispersion near the third-order EP along $k_x$ as well [see Fig. S1B].

Additionally, we also study the energy dispersion near the other third-order EP for $H_{PT}$ with $\epsilon=0.5$. The same features can be derived from the results shown in Figs. S1C and D, where a cube-root energy dispersion away from the third-order EPs located at $(k_x=0.4701, k_y=0.6341)$ is observed experimentally.

As shown in Fig.~2E of the main text, the third-order EPs are destroyed by the symmetry-broken perturbation $\sum_{i=1}^{8}\delta_i\lambda_i$ for $H_{PT}$ with $\epsilon=0.5$. Here $\delta_i$ with $i=1,2,\cdots,8$ are chosen randomly in the region of $[-\pi/20, \pi/20]$. In our experiment, we have $\delta_1=-0.0849$, $\delta_2=0.0531$, $\delta_3=0.0308$, $\delta_4=-0.1390$, $\delta_5=-0.1294$, $\delta_6=-0.0794$, $\delta_7=0.1022$ and $\delta_8=0.1114$.

\begin{figure*}[b!]
  \centering
\includegraphics[width=0.9\textwidth]{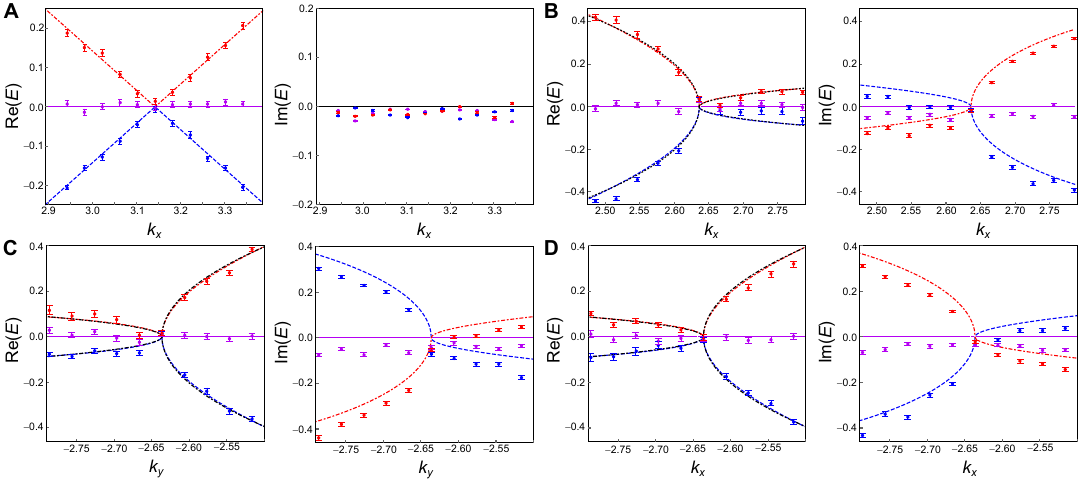}
\caption{{\bf Energy spectra for the P-symmetric three-band model.} Measured (dots) and the corresponding theoretical (lines) eigenenergies of the P-symmetric Hamiltonian $H_{P}$ with fixed $(k_y=0, \epsilon=0)$ for ({\bf A}) and $(k_y=2.6362, \epsilon=0.5)$ for ({\bf B}) as a function of the momentum $k_x$. The energy dispersion near the other third-order EP at $(k_x=-2.6362, k_y=-2.6362)$ are observed by fixing $k_y=-2.6362$ along $k_x$ in ({\bf C}), and fixing $k_x=-2.6362$ along $k_y$ in ({\bf D}). The black dotted lines in the right column of ({\bf B}), ({\bf C}) and ({\bf D}) correspond to the results fitted by $\thicksim k^{1/2}$. Error bars are obtained by assuming Poisson statistics in the photon-number fluctuations, indicating the statistical uncertainty.}
\label{fig:dataS2}
\end{figure*}

\begin{figure*}
  \centering
\includegraphics[width=0.5\textwidth]{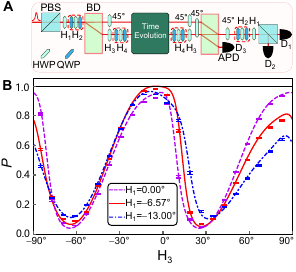}
\caption{{\bf Search the eigenstates of $\tilde{H}_{PT}$.} ({\bf A}) Schematic of the optical circuit used to search the eigenstates. BD: beam displacer; HWP: half-wave plate; QWP: quarter-wave plate; APD: avalanche photodiode; PBS: polarizing beam splitter. ({\bf B}) The measured results of $P$ by fixing $(\text{H}_2=1.82^\circ, \text{H}_4=-11.57^\circ)$ and varying the angles of the HWPs $(\text{H}_1, \text{H}_3)$. Here the evolution is governed by $\tilde{H}_{PT}$ with $(k_x=-0.6529, k_y=-0.5694)$ and $\epsilon=0.5$. Theoretical predictions are represented by colored curves, and experimental results by the corresponding symbols. Error bars are obtained by assuming Poisson statistics in the photon-number fluctuations, indicating the statistical uncertainty.}
\label{fig:dataS3}
\end{figure*}

\subsection{S2 P-symmetry-protected third-order exceptional points.}

We consider the P-symmetry-protected NH Lieb lattice model~\cite{MB21,L89},
\begin{equation}
H_{P}=(1+\cos k_x-i\epsilon)\lambda_1+(1+\cos k_y+i\epsilon)\lambda_6-\sin k_x \lambda_2-\sin k_y \lambda_7
\end{equation}
with eigenenergies $0$ and $E_\pm=\pm\sqrt{2(2+\cos k_x+\cos k_y-i\epsilon\cos k_x+i\epsilon\cos k_y-\epsilon^2)}$. For $\epsilon=0$, $H_{P}$ describes the standard nearest neighbor Hermitian model, which possesses a triple energy degeneracy at $k_x=k_y=\pi$. For $\epsilon \neq 0$, the triple energy degeneracy splits into four third-order EPs, which are located at $(k_x=\pm\arccos(\epsilon^2/2-1), k_y=\pm\arccos(\epsilon^2/2-1))$.

In Figs.~3A and B of the main text, we measure the energy dispersion along $k_y$ by fixing $k_x=\pi$ for $H_P$ with $\epsilon=0$, and $k_x=2.6362$ for $H_P$ with $\epsilon=0.5$. As shown in Figs. S2A and B, by fixing $k_y=\pi$ for $H_P$ with $\epsilon=0$ and $k_y=2.6362$ for $H_P$ with $\epsilon=0.5$, respectively, we sample 11 $k_x$ in our experiment. The measured energy spectrum for $H_P$ with $\epsilon=0$ presents a triple degeneracy with the linear dispersion along $k_x$. For $H_P$ with $\epsilon=0.5$, an anomalous square-root scaling away from the third-order EP along $k_x$ is also observed. In addition, the energy dispersions near the other third-order EP at $(k_x=-2.6362, k_y=-2.6362)$ are shown in Figs. S2C and D. The similar square-root energy dispersion is observed in our experiment.

As shown in Fig.~3D of the main text, the third-order EPs are destroyed by the symmetry-broken perturbation $\sum_{i=1}^{8}\delta_i\lambda_i$ for $H_{P}$ with $\epsilon=0.5$. Here $\delta_i$ with $i=1,2,\cdots,8$ are chosen randomly in the region of $[-\pi/20, \pi/20]$. In our experiment, we have $\delta_1=-0.0569$, $\delta_2=0.0683$, $\delta_3=-0.0988$, $\delta_4=-0.1125$, $\delta_5=0.0595$, $\delta_6=-0.1260$, $\delta_7=-0.1175$ and $\delta_8=0.1544$.

\subsection{S3 Search for the eigenstates.}
As demonstrated in the main text, given an approximate eigenstate, our setup can be used to efficiently estimate the corresponding eigenvalue. However, to calculate the eigenstates by using conventional classical methods requires resources. When the size of the system grows, rapidly increasing of the resources becomes an obstacle. Furthermore, finding the eigenstates of a given Hamiltonian is a fundamental problem and has lots of applications. Here we develop a method of search for the eigenstates of the NH Hamiltonian, which can also be applied to the Hermitian Hamiltonian, apparently.

The search proceeds by preparing the input state of $\ket{\Psi}$. The state is then evolved through a unit-time evolution $\tilde{U}=e^{-i\tilde{H}}$ governed by the NH Hamiltonian $\tilde{H}$. By projecting the evolved state to the initial state, we can obtain the normalized probability of finding the output state unchanged as $P=|\bra{\Psi}\tilde{U}\ket{\Psi}|^2/\bra{\Psi}\tilde{U}^{\dagger}\tilde{U}\ket{\Psi}$. For the evolution operator, it can be rewritten as $\tilde{U}=\sum_ie^{-i\tilde{E}_i}\ketbra{\psi_i}{\chi_i}/\langle{\chi_i}|{\psi_i}\rangle$. Here $\bra{\chi_i}$ and $\ket{\psi_i}$ are the left and right eigenstates of $\tilde{H}$ with the corresponding eigenenergy of $\tilde{E}_i$, which satisfy the relations of
\begin{align*}
\sum_i\frac{\ketbra{\psi_i}{\chi_i}}{\langle{\chi_i}|{\psi_i}\rangle}=\one, \ \langle{\chi_i}|{\chi_i}\rangle=\langle{\psi_i}|{\psi_i}\rangle=1, \ \langle{\chi_i}|{\psi_j}\rangle_{i\neq j}=0.
\end{align*}
For the input state of $\ket{\Psi}$, it can be represented as
\begin{align*}
\ket{\Psi}=\sum_i\frac{c_i\ket{\psi_i}}{\langle{\chi_i}|{\psi_i}\rangle}=\sum_i\frac{b_i\ket{\chi_i}}{\langle{\psi_i}|{\chi_i}\rangle},
\end{align*}
where $c_i=\langle{\chi_i}|{\Psi}\rangle$ and $b_i=\langle{\psi_i}|{\Psi}\rangle$.
Thus, we have
\begin{align*}
P&=\frac{|\bra{\Psi}\tilde{U}\ket{\Psi}|^2}{\bra{\Psi}\tilde{U}^{\dagger}\tilde{U}\ket{\Psi}}
=\sum_{i,j}\frac{c_i^\ast c_j b_i b_j^\ast e^{iE_i^\ast} e^{-iE_j}}
{\langle{\psi_i}|{\chi_i}\rangle\langle{\chi_j}|{\psi_j}\rangle}/
\sum_{i,j}\frac{c_i^\ast c_j e^{iE_i^\ast} e^{-iE_j}\langle{\psi_i}|{\psi_j}\rangle}
{\langle{\psi_i}|{\chi_i}\rangle\langle{\chi_j}|{\psi_j}\rangle}.
\end{align*}
Subtracting the numerator from the denominator, we can obtain
\begin{align*}
&\sum_{i,j}\frac{c_i^\ast c_j e^{iE_i^\ast} e^{-iE_j}\langle{\psi_i}|{\psi_j}\rangle}
{\langle{\psi_i}|{\chi_i}\rangle\langle{\chi_j}|{\psi_j}\rangle}
-\sum_{i,j}\frac{c_i^\ast c_j b_i b_j^\ast e^{iE_i^\ast} e^{-iE_j}}
{\langle{\psi_i}|{\chi_i}\rangle\langle{\chi_j}|{\psi_j}\rangle}
=\sum_{i,j}\frac{c_i^\ast c_j e^{iE_i^\ast} e^{-iE_j}(\langle{\psi_i}|{\psi_j}\rangle- b_i b_j^\ast)}
{\langle{\psi_i}|{\chi_i}\rangle\langle{\chi_j}|{\psi_j}\rangle}\\
&=\sum_{i,j}\frac{c_i^\ast c_j e^{iE_i^\ast} e^{-iE_j}\langle{\psi_i}|(\one-\ketbrad{\Psi})|{\psi_j}\rangle}
{\langle{\psi_i}|{\chi_i}\rangle\langle{\chi_j}|{\psi_j}\rangle}
=\sum_{i,j}\frac{c_i^\ast c_j e^{iE_i^\ast} e^{-iE_j}\sum_{l}\langle{\psi_i}\ketbrad{\Psi^\bot_l}{\psi_j}\rangle}
{\langle{\psi_i}|{\chi_i}\rangle\langle{\chi_j}|{\psi_j}\rangle}\\
&=\sum_{l}\mid\sum_{i}\frac{c_ie^{-iE_i}\langle\Psi^\bot_l|\psi_i\rangle}{\langle\chi_i|\psi_i\rangle}\mid^2\geq0,
\end{align*}
where $\ket{\Psi^\bot_l}$ are the states orthogonal to the input state of $\ket{\Psi}$. Thus, we can obtain $P\leq1$. If and only if $\ket{\Psi}$ is an egigenstate of $\tilde{H}$, the inequality is saturated to an equality with $P=1$. Therefore, $P$ can act as an eigenstate probe, where it equals to the maximum value of $1$ if the input state is an eigenstate of the Hamiltonian.

As proof of principle, we achieve the searching task to prepare the initial state into one of the eigenstates of $\tilde{H}_{PT}$ with $(k_x=-0.6529, k_y=-0.5694)$ and $\epsilon=0.5$. The setup is shown in Fig. S3A, where the input qutrit state is prepared by tuning the setting angles of the half-wave-plates (HWPs) of $\text{H}_{1-4}$. Then, subjecting the state to the evolution governed by the measured Hamiltonian, the probability $P$ can be measured by subjecting the output state to the reverse process of the initial state preparation. We then have $P=N_1/(N_1+N_2+N_3)$, where $N_i$ is the number of heralded clicks at detector $D_i$. By continuously tuning the angles of $\text{H}_{1-4}$ to achieve the maximum $P$, we can prepare the input state as one of the eigenstate of the Hamiltonian, where the searching task can be interpreted as an optimization problem.

By fixing $(\text{H}_2=1.82^\circ, \text{H}_4=-11.57^\circ)$ and varying the angles of the HWPs $(\text{H}_1, \text{H}_3)$ in state preparation and projective measurement, we obtain the maximum measured $P=0.982\pm0.001$ with $(\text{H}_1=-6.57^\circ, \text{H}_3=-5.18^\circ)$. The fidelity between the prepared state and the first eigenstate of $\tilde{H}_{PT}$ is $0.985\pm0.002$.

\subsection{S4 Generalization to symmetry-protected four-band models.}
In our experiment, we first extend the PT-symmetry-protected system to the four-band model governed by the Hamiltonian
\begin{align}
H^\prime_{PT}=k_x\Gamma_1+i\epsilon(\Gamma_2+\Gamma_4+\Gamma_5)+(k_y-i\epsilon)\Gamma_6
-(k_y+i\epsilon)\Gamma_7-\frac{\epsilon}{2}(\Gamma_3+\sqrt{3}\Gamma_8)+\frac{\Gamma_0-\sqrt{6}\Gamma_{15}}{4}
\label{seq:HPT4}
\end{align}
with the symmetry operators $P^\prime=\text{diag}(1, -1, 1, 1)$ and $T^\prime=\text{diag}(-1, 1, i, 1)$. Here $\Gamma_0$ is the $4\times4$ identity matrix and $\Gamma_i$ ($i=1,2,\cdots,15$) denote the Gell-Mann matrices, that span the Lie algebra of the SU(4) group,
\begin{small}
\begin{align}
\Gamma_1&=\begin{pmatrix}
0 & 1 & 0 & 0 \\
1 & 0 & 0 & 0 \\
0 & 0 & 0 & 0 \\
0 & 0 & 0 & 0 \\
\end{pmatrix},
\Gamma_2=\begin{pmatrix}
0 & -i & 0 & 0\\
i & 0 & 0 & 0\\
0 & 0 & 0 & 0\\
0 & 0 & 0 & 0 \\
\end{pmatrix},
\Gamma_3=\begin{pmatrix}
1 & 0 & 0 & 0\\
0 & -1 & 0 & 0\\
0 & 0 & 0 & 0\\
0 & 0 & 0 & 0 \\
\end{pmatrix},
\Gamma_4=\begin{pmatrix}
0 & 0 & 1 & 0\\
0 & 0 & 0 & 0\\
1 & 0 & 0 & 0\\
0 & 0 & 0 & 0 \\
\end{pmatrix}, \\ \nonumber
\Gamma_5&=\begin{pmatrix}
0 & 0 & -i & 0\\
0 & 0 & 0 & 0\\
i & 0 & 0 & 0\\
0 & 0 & 0 & 0 \\
\end{pmatrix},
\Gamma_6=\begin{pmatrix}
0 & 0 & 0 & 0\\
0 & 0 & 1 & 0\\
0 & 1 & 0 & 0\\
0 & 0 & 0 & 0 \\
\end{pmatrix},
\Gamma_7=\begin{pmatrix}
0 & 0 & 0 & 0\\
0 & 0 & -i & 0\\
0 & i & 0 & 0\\
0 & 0 & 0 & 0 \\
\end{pmatrix},
\Gamma_8=\begin{pmatrix}
\frac{1}{\sqrt{3}} & 0 & 0 & 0\\
0 & \frac{1}{\sqrt{3}} & 0 & 0\\
0 & 0 & \frac{-2}{\sqrt{3}} & 0\\
0 & 0 & 0 & 0 \\
\end{pmatrix},\\ \nonumber
\Gamma_9&=\begin{pmatrix}
0 & 0 & 0 & 1\\
0 & 0 & 0 & 0\\
0 & 0 & 0 & 0\\
1 & 0 & 0 & 0 \\
\end{pmatrix},
\Gamma_{10}=\begin{pmatrix}
0 & 0 & 0 & -i\\
0 & 0 & 0 & 0\\
0 & 0 & 0 & 0\\
i & 0 & 0 & 0 \\
\end{pmatrix},
\Gamma_{11}=\begin{pmatrix}
0 & 0 & 0 & 0\\
0 & 0 & 0 & 1\\
0 & 0 & 0 & 0\\
0 & 1 & 0 & 0 \\
\end{pmatrix},
\Gamma_{12}=\begin{pmatrix}
0 & 0 & 0 & 0\\
0 & 0 & 0 & -i\\
0 & 0 & 0 & 0\\
0 & i & 0 & 0 \\
\end{pmatrix},\\ \nonumber
\Gamma_{13}&=\begin{pmatrix}
0 & 0 & 0 & 0\\
0 & 0 & 0 & 0\\
0 & 0 & 0 & 1\\
0 & 0 & 1 & 0 \\
\end{pmatrix},
\Gamma_{14}=\begin{pmatrix}
0 & 0 & 0 & 0\\
0 & 0 & 0 & 0\\
0 & 0 & 0 & -i\\
0 & 0 & i & 0 \\
\end{pmatrix},
\Gamma_{15}=\begin{pmatrix}
\frac{1}{\sqrt{6}} & 0 & 0 & 0\\
0 & \frac{1}{\sqrt{6}} & 0 & 0\\
0 & 0 & \frac{1}{\sqrt{6}} & 0\\
0 & 0 & 0 & -\sqrt{\frac{3}{2}} \\
\end{pmatrix}.
\label{eq:GM}
\end{align}
\end{small}
Using the representation of the P-symmetry operator as $P^\prime$, the P-symmetric non-Hermitian Hamiltonian is restricted to the form of
\begin{equation}
H^\prime_{P}=\frac{(b+d)\Gamma_1+(f+h)\Gamma_6+(\beta+\gamma)\Gamma_{11}}{2}
+\frac{i\left[(b-d)\Gamma_2+(f-h)\Gamma_7+(\beta-\gamma)\Gamma_{12}\right]}{2},
\end{equation}
where $\{b,d,f,h,\beta,\gamma\}$ are arbitrary complex numbers. The specific Hamiltonian satisfies the relations of $\text{det}[H^\prime_{P}]=\text{Tr}(H^{\prime}_{P})=\text{Tr}(H^{\prime3}_{P})=0$ with the corresponding eigenenergies of $\{0, 0, E_\pm^\prime\}$. The two degenerated flat bands with energy zero correspond to the non-defective degeneracies with two distinct eigenstates of $\ket{\psi_1^0}=(-\beta, 0, 0, d)^T/\sqrt{\beta^2+d^2}$ and $\ket{\psi_2^0}=(-f, 0, d, 0)^T/\sqrt{f^2+d^2}$. The two dispersive bands satisfy~\cite{SK22}
\begin{align}
E_{\pm}=\pm\sqrt{\frac{\text{Tr}(H^{\prime2}_{P})}{2}}=\pm\sqrt{bd+fh+\beta\gamma}\nonumber
\end{align}
with the corresponding eigenstates of
\begin{align}
\ket{\psi_3^{+}}=\frac{(b, \sqrt{bd+fh+\beta\gamma}, h, \gamma)^T}{\mathcal{N}}\ \text{and}\ \ket{\psi_4^{-}}=\frac{(b, -\sqrt{bd+fh+\beta\gamma}, h, \gamma)^T}{\mathcal{N}},\nonumber
\end{align}
where $\mathcal{N}=\sqrt{b^2+h^2+\gamma^2+bd+fh+\beta\gamma}$. Thus, we tune two parameters to make $bd+fh+\beta\gamma=0$ satisfied. Then the second-order EPs with $E_\pm=0$ are generated, which also correspond to the four-fold degeneracy points for the NH system. As shown in Figs.~5C and D of the main text, we choose $b=1+e^{ik_x}-i\epsilon$, $d=1+e^{-ik_x}-i\epsilon$, $f=1+e^{ik_y}+i\epsilon$, $h=1+e^{-ik_y}+i\epsilon$ and $\beta=\gamma=\sin k_x$. By setting $\epsilon=0.5$ and tuning $k_x$ and $k_y$, the two dispersive bands touch each other to form the second-order EPs at $\{\pm2.7275, \pm2.7275\}$, where the four-fold degeneracies emerge.

\begin{figure*}
  \centering
\includegraphics[width=0.9\textwidth]{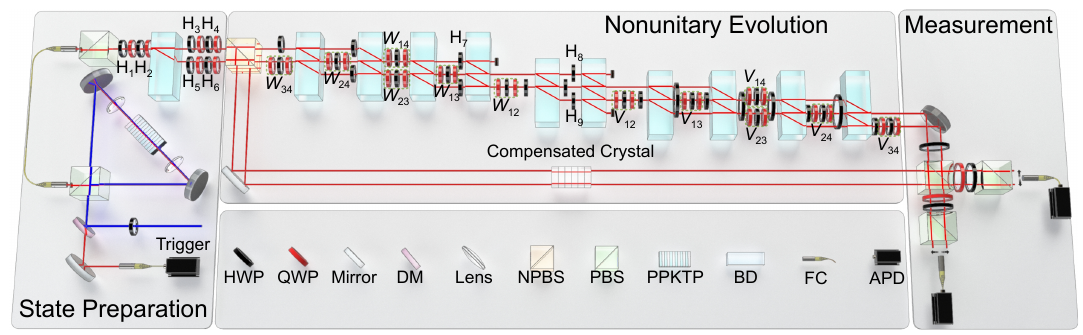}
\caption{{\bf Experimental setup for the four-band models.} DM: dichroic mirror; NPBS: non-polarizing beam splitter; PPKTP: periodically
poled potassium titanyl phosphate crystal; APD: avalanche photodiodes; FC: fiber coupler.}
\label{fig:setup4}
\end{figure*}

As shown in Fig. S4, to construct the energy spectrum of the symmetry-protected four-band models of $H^\prime_{PT}$ and $H^\prime_{P}$ in the main text, the basis states are encoded by the hybrid polarization-spatial modes of the single photons as $\ket{0}\Leftrightarrow\ket{H_1}$, $\ket{1}\Leftrightarrow\ket{V_1}$, $\ket{2}\Leftrightarrow\ket{H_2}$, and $\ket{3}\Leftrightarrow\ket{V_2}$. Here $1,2$ denote the different spatial modes and $H$ ($V$) denotes the horizontal (vertical) polarization of the single photons. The state preparation is achieved by passing the photons through a polarizing beam splitter
(PBS). An adjustable HWP (H$_1$) combined with a sandwich-type set of HWP (H$_2$) and quarter-wave plates (QWPs) with the setting angles of 45$^\circ$ are used to control the amplitude and the relative phases of the photon with different polarizations. After passing through a beam displacer (BD), the vertically polarized photons are transmitted and the horizontal polarized photons go through a $3$-mm lateral displacement into a neighboring mode. Thus, the photons are prepared into two parallel spatial modes---the transmitted first and lateral second modes. By inserting the wave-plates $\text{H}_{1-6}$ into the spatial modes accordingly, we prepare the state of the photons in the eigenstates of the measured Hamiltonian.

After the state preparation, the photons are injected into the $50:50$ non-polarizing beam splitter (NPBS). Thus, extra two paths of the transmitted ($t$) and reflected ($r$) modes are introduced. To approach the nonunitary unit-time evolution governed by the mapped NH Hamiltonian on the transmission, we decompose the nonunitary operation as
\begin{align*}
\tilde{U}_4=e^{-i\tilde{H^\prime}}=V_{34}V_{24}V_{14}V_{23}V_{13}V_{12}D_4W_{12}W_{13}W_{23}W_{14}W_{24}W_{34}.
\end{align*}
Here $\tilde{H^\prime}$ denotes the mapped NH Hamiltonian of $\tilde{H}^\prime_{PT}$ or $\tilde{H}^\prime_{P}$. The operations $V_{ij}$ and $W_{ij}$ are the unitary operations acting on the 2D subspaces of the system with the complementary subspace unchanged, and $D_4$ is a diagonal matrix. As shown in Fig. S4, the unitary operations $V_{ij}$ and $W_{ij}$ can be realized by combining the two acted modes into one spatial mode with different polarizations and applying a $2\times2$ unitary transformation via the set of wave-plates. The diagonal matrix $D_4$ can be realized by introducing the mode-selective losses to the corresponding modes, where the intensity of the losses can be controlled by HWPs of H$_{7-9}$.

After recombining the photons in the $t$- and $r$- modes, the projective measurements are performed on the polarization of the photons in different spatial modes with the bases of $\{\ket{\pm}=\left(\ket{H}\pm\ket{V}\right)/\sqrt{2}, \ket{R}=\left(\ket{H}-i\ket{V}\right)/\sqrt{2}\}$. The coincidences for the projective measurements are counted as $\{N_i^{\pm}, N_i^{R}\}$. Here $i\in\{t_1^\prime, r_1^\prime, t_2^\prime, r_2^\prime\}$ denotes the transmitted ($t^\prime$) or reflected ($r^\prime$) path following the second PBS in the first or second spatial modes of the $r$-path photons. Accordingly, we can obtain the eigenenergy of $\tilde{H^\prime}$ as the complex phase shift between $t$ and $r$ paths, i.e.,
\begin{align*}
\xi^\prime=e^{-i\tilde{E}^\prime_j}=\bra{\psi_j}\tilde{U}\ket{\psi_j}=&\frac{\sum_i\left(N^+_{i}-N^-_{i}\right)+i\sum_i\left(N^+_{i}+N^-_{i}-2N^R_{i}\right)}{N_\text{tot}}\\ \nonumber
&-\frac{2i\left(N^+_{r_1^\prime}+N^-_{r_1^\prime}-2N^R_{r_1^\prime}\right)+2i\left(N^+_{r_2^\prime}+N^-_{r_2^\prime}-2N^R_{r_2^\prime}\right)}{N_\text{tot}}\nonumber,
\end{align*}
where $N_\text{tot}$ denotes the number of the input photons to achieve the state preparation. The eigenenergy $E^\prime_j$ of $H^\prime_{PT}$ or $H^\prime_{P}$ can thus be obtained through the relation of $E^\prime_j=i\ln\xi^\prime+i\ln\sqrt{\Lambda^\prime}$, where $\Lambda^\prime=\max_{\mathbf{k}}|\zeta_{\mathbf{k}}|$ and $\zeta_{\mathbf{k}}$ is the eigenvalue of $e^{-iH^\prime_{PT}}e^{-iH^{\prime\dagger}_{PT}}$ or $e^{-iH^\prime_{P}}e^{-iH^{\prime\dagger}_{P}}$.

\subsection{S5 Generalization to arbitrary models.}
Our setup can be efficiently generalized to construct the energy spectrum of arbitrary NH Hamiltonians, by taking advantage of the extendable degrees of freedom of photons and specially designed interferometric network.

As shown in Fig.~1 of the main text, the process starts by preparing the initial state in one of the eigenstates of the corresponding Hamiltonian.
For an $N$-dimensional qudit state, the basis states can be encoded as~\cite{WKZ17}
\begin{align*}
\left(\ket{0}, \ket{1}\right),\left(\ket{2}, \ket{3}\right),\cdots,\left(\ket{N-1}, \ket{N}\right)\Longleftrightarrow
\left(\ket{H_1}, \ket{V_1}\right),\left(\ket{H_2}, \ket{V_2}\right),\cdots,\left(\ket{H_n}, \ket{V_n}\right)\nonumber
\end{align*}
for an even $N$, where $n=N/2$ denotes the number of the spatial modes and $H$ ($V$) denotes the horizontal (vertical) polarization of the single photons. Whereas, if $N$ is odd, the basis states are encoded as
\begin{align*}
\left(\ket{0}\right),\left(\ket{1}, \ket{2}\right),\cdots,\left(\ket{N-1}, \ket{N}\right)\Longleftrightarrow
\left(\ket{H_1}\right),\left(\ket{H_2}, \ket{V_2}\right),\cdots,\left(\ket{H_n}, \ket{V_n}\right)\nonumber
\end{align*}
with $n=(N+1)/2$.

To extend the proposed method to achieve the preparation of the initial state, it is convenient to expand the spatial modes by additional BDs and tune the parameters by using the wave-plates. Actually, for an $N$-dimensional pure states, it can be prepared by using $(N-1)/2$ BDs when $N$ is an odd number and $N/2-1$ for an even number. All the proportions of each mode and the relative phases can be tuned by varying the angles of $N-1$ HWPs and $N-1$ sandwich-type sets of QWP-HWP-QWP.

After passing the initial state through the nonunitary unit-time evolution governed by the mapped NH Hamiltonian, the corresponding eigenenergies will be introduced as the complex phase shift to the evolved state. The decomposition method presented in the main text to achieve the nonunitary dynamics of $\tilde{U}$ can also be generalized to any $N\times N$ passive nonunitary operation as $\tilde{U}_N=\prod_{n=N}^{n=1}\prod_{m=n-1}^{m=1}V_{mn}D_N\prod_{i=1}^{i=N}\prod_{j=i+1}^{j=N}W_{ij}$.
Thus, the simple building blocks of BDs and wave plates to approach the corresponding decomposed operations provide a recipe for an implementation of arbitrary $\tilde{U}_N$, experimentally. The maximum numbers of BDs needed to build all the $N$-dimensional unitary operators of $W_{ij}$ and $V_{mn}$ are both $2N-4$ when $N$ is an even number and $2N-3$ for an odd number. The number of BDs to achieve $D_N$ is $N-1$. Thus, The maximum number of BDs to approach this process is only linear increasing with $N$. Besides, to fully control the parameters in the nonunitary operation, the maximum number of set of wave plates to approach the decomposed unitary operations is $(N-1)N$, and $N-1$ HWPs to realize $D_N$. At last, by performing interferometric measurements on the photons, the complex eigenenergies for each mode in reciprocal space can be constructed experimentally.

\end{document}